\documentclass[prl,preprint,superscriptaddress]{revtex4-2}
\usepackage[margin=1in]{geometry}
\usepackage[mathlines]{lineno}
\usepackage{setspace}
\usepackage[utf8]{inputenc}
\usepackage{graphicx}
\usepackage{amsmath}
\usepackage{xcolor}
\usepackage{soul}
\usepackage{mathtools}

\newcommand{\Min}{\mathrm{min}}
\newcommand{\NiP}{\text{4-NiP}}
\newcommand{\AmP}{\text{4-AmP}}
\newcommand{\BH}{\text{NaBH}_4}
\newcommand{\Hy}{\text{H}_2}
\newcommand{\SiO}{\mathrm{SiO}_2}
\newcommand{\mg}{\mathrm{mg}}
\newcommand{\g}{\mathrm{g}}
\newcommand{\CaCl}{\mathrm{CaCl}_2}
\newcommand{\Pt}{\text{Pt}}

\begin{document}

\title{Effects of Hydrogen Transport on the Kinetic Regimes of 4-Nitrophenol Reduction by Sodium Borohydride}

\author{Tatiana Nizkaia}

\affiliation{Helmholtz-Institute Erlangen-N\"urnberg for Renewable Energy (IET-2), Forschungszentrum J\"ulich, Cauerstra\ss e 1, 91058 Erlangen, Germany}

\author{Philipp Groppe}
\affiliation{Department of Chemistry and Pharmacy, Friedrich-Alexander-Universit\"at Erlangen-N\"urnberg (FAU), Egerlandstra\ss e 1, 91058 Erlangen, Germany}
\author{Valentin M\"uller}
\affiliation{Department of Chemistry and Pharmacy, Friedrich-Alexander-Universit\"at Erlangen-N\"urnberg (FAU), Egerlandstra\ss e 1, 91058 Erlangen, Germany}
\author{Jens Harting}
\affiliation{Helmholtz-Institute Erlangen-N\"urnberg for Renewable Energy (IET-2), Forschungszentrum J\"ulich, Cauerstra\ss e 1, 91058 Erlangen, Germany}
\affiliation{Department of Chemical and Biological Engineering and Department of Physics, Friedrich-Alexander-Universit\"at Erlangen-N\"urnberg, Cauerstraße 1, 91058 Erlangen, Germany}
\author{Susanne Wintzheimer}
\affiliation{Department of Chemistry and Pharmacy, Friedrich-Alexander-Universit\"at Erlangen-N\"urnberg (FAU), Egerlandstra\ss e 1, 91058 Erlangen, Germany}
\affiliation{Fraunhofer-Institute for Silicate Research ISC, Neunerplatz 2, 97082 W\"urzburg, Germany}
\author{Paolo Malgaretti}
\affiliation{Helmholtz-Institute Erlangen-N\"urnberg for Renewable Energy (IET-2), Forschungszentrum J\"ulich, Cauerstra\ss e 1, 91058 Erlangen, Germany}

\onehalfspacing
\begin{abstract}
The reduction of 4-nitrophenol (4-NiP) by sodium borohydride is widely used to benchmark heterogeneous catalysts and is commonly simplified as a pseudo-first-order reaction, characterized by a single reaction rate constant. In reality, this reaction is more complex, as it is accompanied by hydrolysis of borohydride and concurrent hydrogenation of 4-NiP by produced hydrogen. This makes the local hydrogen concentration at catalytic sites an important, and so far overlooked, factor in shaping the apparent catalytic activity of heterogeneous catalysts. Re-examining benchmarking experiments on Pt–$\SiO$ supraparticles with different pore structures, we attribute contrasting kinetic behavior to distinct regimes of hydrogen transport: diffusive transport sustains high local concentrations of hydrogen and pseudo-first-order kinetics of 4-NiP hydrogenation, while bubble-mediated transport causes hydrogen loss, deviations from pseudo-first-order regime and incomplete conversion of 4-NiP. We propose a kinetic model that captures this transition and enables consistent interpretation of experimental data. More broadly, our analysis shows that apparent differences in activity observed in benchmarking experiments, that use 4-NiP reduction by borohydride as a test reaction, can arise from hydrogen transport rather than intrinsic properties of the catalyst. This highlights the need to account for the hydrogen transport regime (bubbling/non-bubbling), when comparing catalyst performance across different experiments.
\end{abstract}

\maketitle

\section*{Introduction}
The reduction of 4-Nitrophenol (4-NiP) by  sodium borohydride (\(\text{NaBH}_4\)) in aqueous solutions is widely recognized as a benchmark reaction for evaluating the efficiency of heterogeneous catalysts~\cite{herves2012catalysis,goepel2014hydrogenation,aditya2015nitroarene,das2022advances}. It has been first described over 20 years ago \cite{pradhan2001catalytic,ghosh2004bimetallic} and quickly gained popularity due to its operational simplicity and the convenience of \textit{in situ} monitoring using UV-visible spectroscopy. 
Traditionally, the reaction kinetics have been described as pseudo-first-order, characterized by a single reaction rate constant which depends on initial concentrations of $\NiP$ and $\BH$~\cite{wunder2010kinetic,thawarkar2018kinetic,ayad2020kinetic}.

However, recent studies have shown that this reaction is more complex than previously thought. In fact, the catalysts used for the reaction also promote the hydrolysis of borohydride, leading to the formation of molecular hydrogen~\cite{varshney2023competition,grzeschik2020overlooked,serra2020simple}. Since this process is much faster than 4-NiP reduction, experiments are typically carried out with a large excess of $\BH$ to ensure complete conversion of 4-NiP and to maintain pseudo-first-order kinetics~\cite{herves2012catalysis}. Also, some metals can catalyse the hydrogenation of 4-NiP by dissolved hydrogen~\cite{vaidya2003synthesis,paun2016flow,grzeschik2020overlooked}. In this case, some of the hydrogen produced by hydrolysis can be used for direct 4-NiP hydrogenation \cite{grzeschik2020overlooked}, while the rest is transported away from the catalytic region in the form of bubbles or via diffusion, and eventually leaves the reactor. The 4-NiP to 4-AmP conversion rate in this case depends on the balance between hydrogen production and the rate at which it leaves the system, which ultimately depends on the onset of bubbling~\cite{oehmichen2010influence,solymosi2022nucleation}. Accordingly, as it has recently been discussed for other catalytic systems~\cite{heenen2024mesoscopic}, transport of hydrogen becomes a crucial factor in determining the catalyst efficiency.

\begin{figure}[h!]
    \centering
    \includegraphics[width=0.65\textwidth]{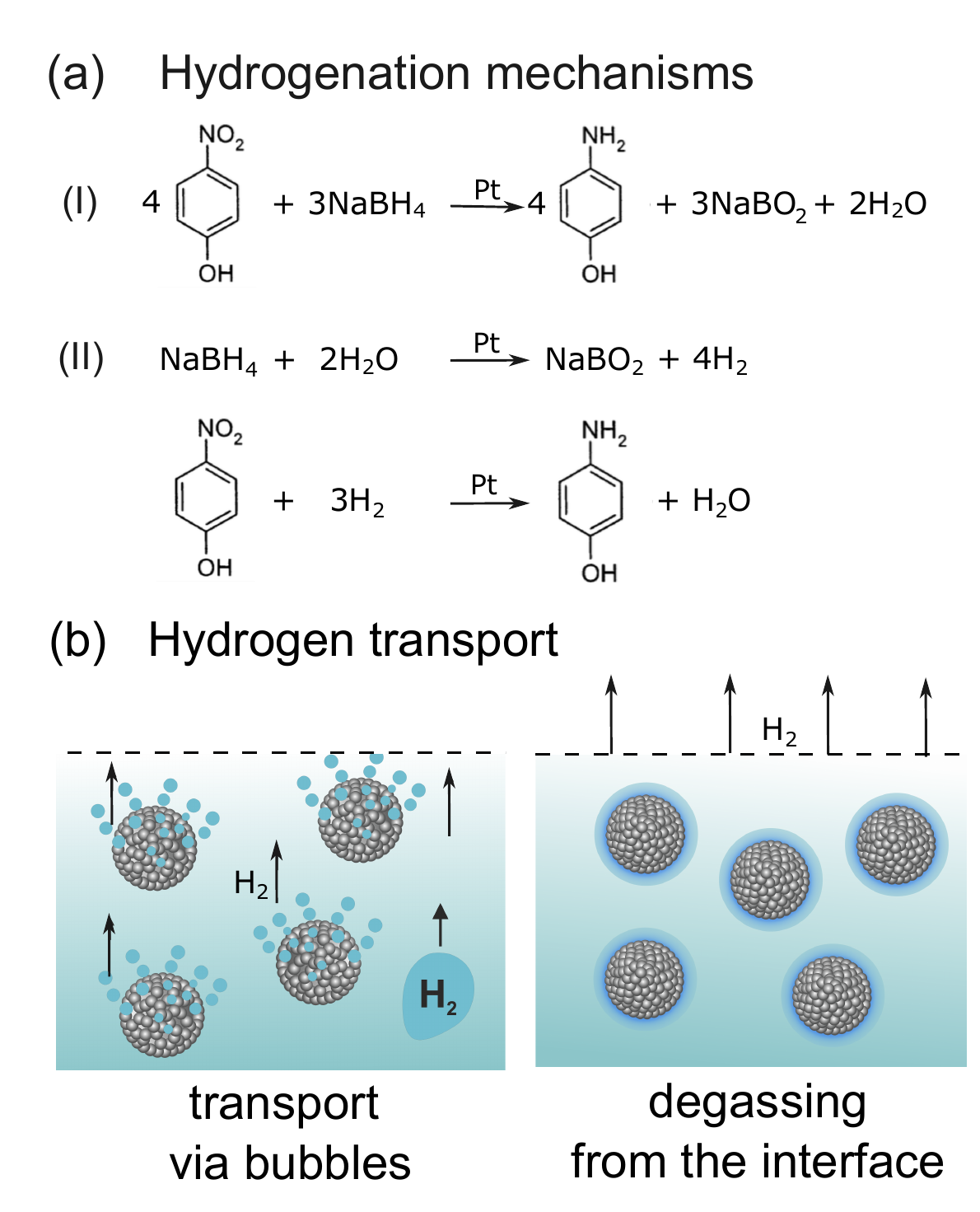}
    \caption{(a) Reaction scheme of 4-NiP hydrogenation, depicting different mechanisms: (I) transfer hydrogenation by borohydride and (II) hydrolysis of borohydride and hydrogenation by dissolved hydrogen.
    (b) Sketch of hydrogen transport mechanisms: bubbling vs. degassing from the liquid/gas interface.}
    \label{fig:sketch}
\end{figure}

While deviations of 4-NiP reduction kinetics from classical models have been observed previously~\cite{gu2014kinetic,gu2015kinetic,li2013revisiting} and hydrogenation via dissolved $\Hy$  has been acknowledged \cite{grzeschik2020overlooked}, the transport of hydrogen has never been considered as a factor that can affect the reaction kinetics.
In this study, we address its relevance by re-examining experimental data on the reduction of 4-NiP using $\Pt-\SiO$ catalytic supraparticles (SPs) fabricated via spray-drying techniques~\cite{groppe2024catalyst,groppe2025atomic}. These SPs offer tunable porosity at a fixed catalyst loading, enabling investigations into how pore size and distribution of the catalyst affect the reaction kinetics. While assessing the catalytic activity of the obtained supraparticles using 4-NiP hydrogenation as a test reaction, we observed puzzling deviations from the expected reaction kinetics, particularly in samples that exhibit bubble formation. 
To address these deviations, we develop a kinetic model that includes 4-NiP reduction by both sodium borohydride and dissolved $\Hy$, and takes into account the transport of hydrogen. The model reveals two distinct regimes: initially, reduction by borohydride dominates; at later stages, once borohydride is depleted, the main reduction agent is dissolved $\Hy$. We use our model to analyze experimental data and find evidence for $\Hy$-mediated hydrogenation at long times and for the influence of hydrogen transport on the apparent reaction kinetics.

\section*{Kinetic model of the two reduction mechanisms}

The classical scheme of 4-NiP reduction~\cite{saha2010photochemical,wunder2010kinetic,gu2014kinetic} describes a binary reaction at the surface of a catalytic metal nanoparticle (M-NP) as \cite{varshney2023competition}:
\begin{equation}
4\text{HO}(\text{C}_6\text{H}_4)\text{NO}_2 + \text{3BH}_4^{-} \xrightarrow{\text{M-NP}}4\text{HO}(\text{C}_6\text{H}_4)\text{NH}_2  + 3\text{BO}_2^-+ 2\text{H}_2\text{O}\label{react:trans_hydr}
\end{equation}
This process is accompanied by catalytic hydrolysis of borohydride~\cite{varshney2023competition},
\begin{equation}
\text{BH}_4^- + 4\,\text{H}_2\text{O} \xrightarrow{\text{M-NP}} \text{B(OH)}_4^- + 4\,\text{H}_2, \label{react:hydrolysis}
\end{equation}
which is typically much faster; hence, an excess of borohydride is required for efficient 4-NiP reduction~\cite{aditya2015nitroarene}. Meanwhile, some metals (in particular Pt and Pd) also catalyse a direct hydrogenation reaction with $\Hy$ produced by hydrolysis~\cite{vaidya2003synthesis,grzeschik2020overlooked,paun2016flow}:
\begin{equation}
   \text{HO}(\text{C}_6\text{H}_4)\text{NO}_2 + 3\,\text{H}_2 \xrightarrow{\text{M-NP}} \text{HO}(\text{C}_6\text{H}_4)\text{NH}_2 + 2\,\text{H}_2\text{O},\label{react:hydrogenation}
\end{equation}
Experimental characterization of this pathway under high $\Hy$ pressure shows its independence on pH and pseudo-zeroth-order kinetics with respect to 4-NiP, up to very small 4-NiP concentrations (see~\cite{vaidya2003synthesis} and the Supplementary material in \cite{grzeschik2020overlooked}).

Therefore, two mechanisms can coexist: reduction by sodium borohydride (Eq.~\ref{react:trans_hydr}) and a two-step route via hydrolysis and hydrogenation by dissolved $\Hy$ (Eqs.~\ref{react:hydrolysis}--\ref{react:hydrogenation}). When $\Hy$ is produced, it can either leave the system or be used for 4-NiP hydrogenation. The fraction used is controlled by the competition between the hydrolysis rate and $\Hy$ removal. Accordingly, fast transport reduces the local $\Hy$ concentration and thus the overall conversion rate of 4-NiP. Prior to borohydride depletion, both mechanisms are likely operational, leading to complex reaction kinetics.
To capture both chemical pathways together with the transport of hydrogen, we use a simplified kinetic model:
\begin{subequations}\label{eq:model-fran}
\begin{align}
\dfrac{dC_{\NiP}}{dt}&= - k_A C_{\Hy}-4k_{AB} C_{\NiP} C_{\BH},\label{eq:frank_nip}\\
\dfrac{dC_{\BH}}{dt}&= -k_B C_{\BH}-3k_{AB} C_{\NiP} C_{\BH},\label{eq:frank_bh}\\
\dfrac{dC_{\Hy}}{dt}&= 4k_B C_{\BH}-3k_A C_{\Hy}-\alpha C_{\Hy}.\label{eq:frank_h2}
\end{align}
\end{subequations}
Here, $C_{\BH}$, $C_{\NiP}$, $C_{H_2}$ are concentrations of the respective species, $k_{AB}$ and $k_A$ are effective rate constants of 4-NiP reduction by borohydride and $\Hy$-mediated hydrogenation, $k_B$ is the hydrolysis rate, and $\alpha$ is the $\Hy$ transport rate. Note that these rate coefficients are effective, i.e., they depend on catalyst loading, distribution of catalyst inside the supraparticles, and adsorption kinetics. We remind that we use here a pseudo zeroth order kinetic model for $H_2$-mediated hydrogenation \cite{vaidya2003synthesis}, which is supposed to break down at very small concentrations of $\NiP$. Our model also neglects the back reaction of $\AmP$ with oxygen \cite{menumerov2016catalytic}, which starts to play a role when oxygen adsorbed from the atmosphere dominates over hydrogen produced by hydrolysis: this can happen in the first seconds of the reaction ("induction period") or when the volume of the reaction mixture is very small, resulting in high surface-to-volume ratio. Assumptions and limitations of this model are discussed in more detail in Supplementary Materials.

Incorporating a transport term, $\alpha C_{\Hy}$, into the kinetic model (see Eq.\eqref{eq:frank_h2}) allows us to account for the rate at which hydrogen escapes from the solution. Notably, when hydrogen is released in the form of bubbles, its evacuation rate is significantly higher than in the absence of bubbling, where hydrogen loss is limited to degassing from the liquid surface. To capture changes in the transport mechanism over the course of the reaction, we introduce a simplified, piece-wise constant model for the transport coefficient:
\begin{align}
    {\alpha}(t)=\left\{\begin{array}{cc}
         {\alpha}_s & t<t_{bub}\\
         {\alpha}_l & t\geq t_{bub}, 
    \end{array}\right.\label{eq:alpha_t}
\end{align}
where $\alpha_s$ corresponds to bubble-mediated $H_2$ transport, $\alpha_l$ is the transport coefficient in the absence of bubbles, and $t_{bub}$ is the time at which bubbling stops, which can be determined by visual observation. 

The transport coefficient in the absence of bubbling $\alpha_l$ can be estimated as the degassing rate from a stirred beaker: $\alpha_l = k_L a$, where $a = A/V$ is the gas--liquid interfacial area per unit volume, and $k_L$ is the liquid-side mass-transfer coefficient which captures the effective rate of transport of chemical species to the interface \cite{danckwerts1970gas}.
Since $\alpha_l$ is proportional to surface-to-volume ratio, it is not an intrinsic property and depends on the volume of the reaction mixture and on the shape of the reservoir in which the reaction takes place. Accordingly, in order to compare different experimental observations it is important to know (and hence to report) these data.

On the other hand, bubble-mediated transport coefficient $\alpha_s$ should be regarded as an empirical parameter: the flux of hydrogen in the bubbling regime depends strongly on the conditions of bubble growth and detachment, which are sensitive to the porous structure and wetting properties of the catalyst support, stirring rate, etc.

\section*{Experimental Methods}
\subsection*{Synthesis of catalytic supraparticles}
In this work we analyse the kinetics of 4-NiP reduction on three different types of catalytic supraparticles (Typea A,B and C):

\textbf{Type A and Type B Pt--$\mathrm{SiO_2}$ Supraparticles.}
Catalyst Types A and B correspond to the spray-dried $\Pt$--$\mathrm{SiO_2}$ supraparticles previously reported by Groppe \textit{et~al.}~\cite{groppe2024catalyst}. Their synthesis and structural characterization are described in detail in that publication. Briefly, colloidal suspensions containing $4$~nm Pt nanoparticles and $\SiO$ nanoparticles of different sizes in a $\CaCl$ solution were spray-dried to form hierarchical supraparticles with tunable pore frameworks. Type~A supraparticles were prepared from 182~nm $\mathrm{SiO_2}$ nanoparticles with varying $\mathrm{CaCl_2}$ concentration (0.025-0.075~mmol~$\mathrm{Ca^{2+}}/\g_\mathrm{SiO_2}$), producing large interstitial pores ($>$40~nm).  Type~B supraparticles were made from 19~nm $\mathrm{SiO_2}$ nanoparticles with varying $\mathrm{CaCl_2}$ concentration (0.1-0.25~mmol~$\mathrm{Ca^{2+}}$~g$^{-1}$~$\mathrm{SiO_2}$) and feature smaller pores (8--19~nm). The Pt loading, measured via
inductively coupled plasma atomic emission spectroscopy (ICP AES), for Type A and Type B particles was similar: $0.9$ and $0.94\, \mg_{\Pt}/\g_{\SiO}$ respectively \cite{groppe2024catalyst}. All structural and compositional data are reproduced from ref.~\cite{groppe2024catalyst}, and no additional synthesis or measurements were performed in the present work.

\textbf{Type C $\Pt$--$\mathrm{SiO_2}$ Supraparticles.}
Catalytic supraparticles of Type~C were newly synthesized following the protocol reported by Groppe \textit{et~al.}~\cite{groppe2025atomic}. $\mathrm{SiO_2}$ supraparticles were produced by spray-drying an aqueous dispersion of 8~nm $\mathrm{SiO_2}$ nanoparticles without $\mathrm{CaCl_2}$ to yield compact structures with pore diameters of $2-10$~nm. Platinum was subsequently deposited by atomic layer deposition (ALD) conducted on a
GEMStar-6 XT ALD reactor from Arradiance using trimethyl(methylcyclopentadienyl)platinum(IV) and ozone as precursors. Two catalyst batches (15 and~30~ALD cycles) were prepared. All ALD and spray-drying parameters were reproduced from ref.~\cite{groppe2025atomic}.

\textbf{Catalyst stability.} 
For particles of Type A and B, stability of the catalyst has been studied in \cite{groppe2024catalyst}. SEM and $N_2$-sorption measurements of the materials after reaction showed no detectable structural degradation (see Fig. S43–S44, Table S9 in \cite{groppe2024catalyst}), and cycle stability tests confirmed reproducible activity over two reaction cycles (see Fig. S45 in \cite{groppe2024catalyst}). Moreover, cycle stability tests showed no loss of activity over two consecutive catalytic runs (Figure S45). A slight activity increase after the second cycle was attributed to minor disintegration of individual Pt nanoparticles, which increases the accessible active surface area. Therefore, catalyst deactivation can be ruled out as the origin of the deviations from pseudo-first-order kinetics observed in this work. These results indicate that the supraparticle catalysts remain stable under the reaction conditions used in this work.

\textbf{Reproducibility.}  The spray-drying fabrication method yields highly reproducible pore structures of the silica supraparticles across independent synthesis batches, confirmed by structural analysis (see Fig. S16 and Table S6 in the Supporting Information of \cite{groppe2025atomic}). 
The ALD-based Pt deposition can show slight batch-to-batch variations regarding the total amount of deposited Pt. However, by quantifying the Pt content using ICP (inductively coupled plasma) measurements, its effects on the apparent fluctuations in the rate constant can be taken into account.

\subsection*{4-NiP reduction experiments}
The time-resolved 4-NiP concentration curves for catalysts of Type A and B were taken directly from ref.~\cite{groppe2024catalyst} and reanalyzed using the kinetic framework introduced in this study. No new catalytic experiments were carried out for Types~A or~B. For Type C catalysts, experiments were performed under identical conditions to those used in Groppe et al.\cite{groppe2024catalyst}: $C_{\NiP}=6.5 \cdot 10^{-5} M$, $C_{\BH}=0.1 M$ in aqueous solution at $T=25^\circ C$ and $\text{pH}=10.4-10.5$ set by NaOH addition. The reaction was performed in a 25~mL beaker (30~mm diameter), agitated with a magnetic stirrer at 200 rpm. The reaction was monitored by UV–Vis spectroscopy of filtered samples of the reaction mixture, extracted by 2 ml portions at 2 min intervals, starting from the initial volume $V=20$ ml \cite{groppe2024catalyst,groppe2025atomic}. The decrease in 400 nm absorbance peak corresponding to 4-NiP was used to determine 4-NiP concentration as a function of time:
\begin{equation}
c_{\NiP}(t) = \frac{C_{\NiP}(t)}{C_{\NiP}(0)} = \frac{A(t)}{A(0)},\label{eq:c(t)}
\end{equation}
where $A(t)$ is the amplitude of the peak at time $t$ \cite{groppe2024catalyst}. The bubbling time was assessed by visual inspection of the video recordings of the experiments.

To assess the importance of $H_2$-mediated reduction, we also performed the experiments for Type C particles using gaseous hydrogen as reducing agent. Experiments were performed under the same conditions as described above for 15-cycle Pt ALD on 8~nm SiO\(_2\) spherical particles. Reactions were carried out at 25~$^\circ$C in a 25~mL beaker  containing 20~mL of 4-nitrophenol solution ($7.5^{-5} M$). A 2~mL aliquot was withdrawn at $t=0$~min, after which H\(_2\) was introduced via a needle at a flow rate of 10~mL~min$^{-1}$. Successively 2 ml portions have been taken out every 2 minutes to perform UV-Vis spectroscopy.

\section*{Analysis of experimental data }

We revisit here the results of 4-NiP reduction experiments on spray-dried composite $\Pt-\SiO$ supraparticles described in~\cite{groppe2024catalyst} (Type A and B) and complement them with new measurements for superparticles obtained by spray-drying a dispersion of passive $\SiO$ nanoparticles with subsequent incorporation of Pt using the ALD method~\cite{groppe2025atomic}:

\textbf{Type A: 182 nm $\SiO$ + 4 nm $\Pt$ + $\CaCl$.} Large pores (all $>40$ nm), agglomerates of Pt nanoparticles inside the supraparticles and on the surface; strong bubbling in the first 5--8 min. Dataset from ref.~\cite{groppe2024catalyst}.

\textbf{Type B: 19 nm $\SiO$ + 4 nm $\Pt$ + $\CaCl$.} Tunable pore size distribution (8--19 nm) with all pores $<40$ nm; aggregates of Pt particles inside the supraparticles; no bubbling observed. Dataset from ref.~\cite{groppe2024catalyst}.

\textbf{Type C: 8 nm $\SiO$ with $\Pt$ by ALD}~\cite{groppe2025atomic}. Very small pores (4 nm); small clusters of Pt atoms in a thin layer near the surface of the supraparticle. 4-NiP reduction experiments performed for supraparticles with 15 and 30 ALD cycles show active bubble formation in the initial stage of the reaction: 5--7 min (30 cycles) and 7--9 min (15 cycles). For particles fabricated with 15 ALD cycles an additional experiment was performed with $H_2$ as a reducing agent, which was bubbled through the $\NiP$ solution at a constant rate of $10 \mathrm{mL} /\mathrm{min}$.

\begin{figure*}[]
    \centering
    \includegraphics[width=\textwidth]{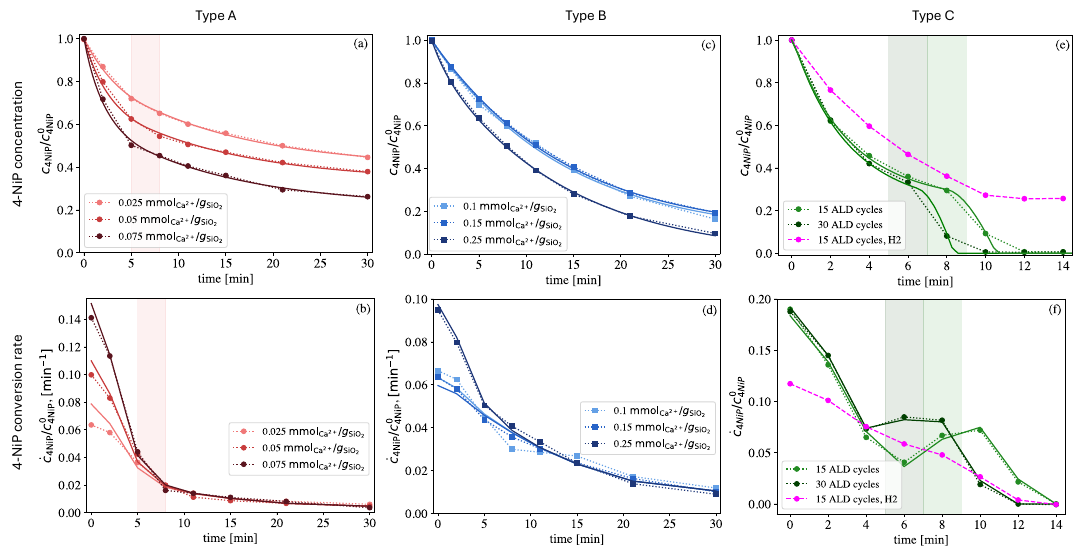}   \caption{Experimental data: time evolution of normalized 4-NiP concentration and reaction rate for Types A (a,b), B (c,d), and C (e,f). Dotted curves with symbols: experimental data; solid curves: model fits to Eqs.~(\ref{eq:frank_nip}--\ref{eq:frank_h2}). The dashed line in (e,f) corresponds to the experiment on 4-NiP reduction with $H_2$. Shaded areas: time range within which bubbling stops.} 
    \label{fig:logs}
\end{figure*}

The evolution of the $\NiP$ concentration for supraparticles of Types A, B, and C is shown in Fig.~\ref{fig:logs}a,c,e (experimental data are shown as symbols connected by dotted lines). While the initial reaction rates for Types A and B are similar , Type A particles show weaker activity at longer times and lower 4-NiP conversion after 30 min (see Fig.~\ref{fig:logs}b,d). Type C particles show higher initial reaction rates compared to particles of Type A and B, and experience an unexpected surge of activity a few minutes into the reaction (see Fig.~\ref{fig:logs}e,f). Type C particles also exhibit high catalytic activity for the $\Hy$ reduction of 4-nitrophenol (magenta curves in Fig.~\ref{fig:logs}e,f), confirming the viability of the $\Hy$-mediated pathway.

\begin{figure}[h]
    \includegraphics[width=0.45\textwidth]{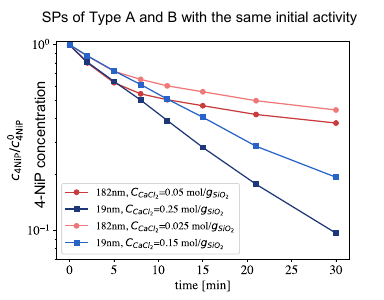}
   \caption{Semi-logarithmic plot of normalized 4-NiP concentration for Type A and Type B supraparticles with the same initial activity.}
    \label{fig:compare}
\end{figure}

We start by comparing the activity of Type A and Type B supraparticles, which all have the same $\Pt$ loading. Interestingly, as shown in Fig.~\ref{fig:compare}, some pairs of Type A/Type B supraparticles exhibit identical catalytic activity during the first 5\,min of the reaction, but diverge markedly at longer times. Moreover, Fig.~\ref{fig:compare} shows that the concentration profiles of Type B supraparticles remain linear on a semilogarithmic scale, consistent with pseudo-first-order kinetics, whereas Type A profiles deviate after 5–8 min, indicating a transition to a different kinetic regime. To pinpoint this transition, we extract the time evolution of the 4-NiP reduction rate by applying finite-difference differentiation to the measured concentration profile in Fig.\ref{fig:logs}a,c,e. As shown in Fig.\ref{fig:logs}b,d at short times ($t<5$ min), the reaction rates differ between different Type A samples, but at longer times ($t>5$ min), all the rate curves collapse. Notably, while the 4-NiP concentration at $t=5$ min varies significantly (Fig.~\ref{fig:logs}a), the rates at and beyond this point are nearly identical (Fig.~\ref{fig:logs}b). This behavior is consistent with the pseudo\text{-}zeroth\text{-}order kinetics with respect to 4-NiP, characteristic of $\Hy$-mediated hydrogenation reported in Refs.~\cite{vaidya2003synthesis,grzeschik2020overlooked}. This can happen if most of the borohydride is hydrolysed in the first few minutes of the reaction, after which  hydrogenation by dissolved $\Hy$ dominates. This suggestion is also consistent with the fact that production of $\Hy$ bubbles was observed only in the first  5–8 min of the reaction~\cite{groppe2024catalyst}. 

\begin{figure*}[t!]
    \centering
     \includegraphics[width=0.99\textwidth]{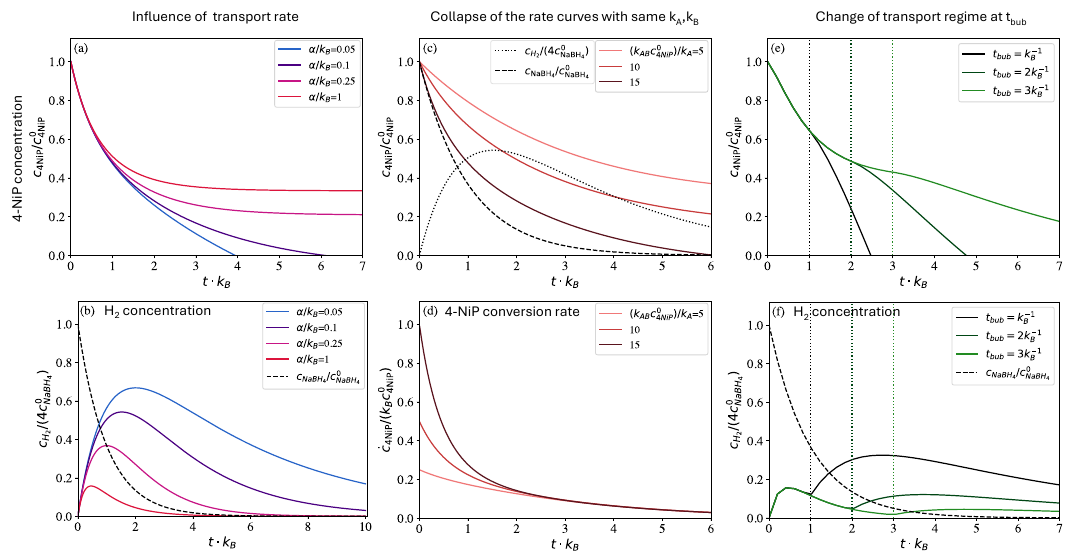}
   \caption{Theoretical results: calculations using model Eqs.(\ref{eq:frank_bh}-\ref{eq:frank_h2}) reproducing typical kinetic features observed in experiments. }
    \label{fig:theory}
\end{figure*}
To analyze the data, we use a simple kinetic model Eqs.~(\ref{eq:frank_nip}--\ref{eq:frank_h2}), which describes two pathways of 4-NiP reduction and takes into account hydrogen transport (see Supplementary material for details). Before fitting the model to experimental data, we use it to rationalize the observed kinetic features on a qualitative level. 

First, our model predicts that for systems with identical reaction rates ($k_{AB}, k_A, k_B$, see the reaction scheme in  Fig.\ref{fig:sketch} and Eqs.(\ref{eq:frank_nip}-\ref{eq:frank_h2})) and different transport rates $\alpha$ the 4-NiP concentration curves collapse at short times, then diverge at long times to reach different conversion levels (Fig.~\ref{fig:theory}a,b). Indeed, at short times, the reduction by sodium borohydride dominates and the reaction rate is governed by $\BH$ concentration (dashed curve in Fig.~\ref{fig:theory}a,b), which is the same for all the systems; at long times, reduction proceeds via dissolved $\Hy$, whose concentration depends on the transport rate. This is exactly what we observe for supraparticles of Types A and B with matched initial rates (see Fig.~\ref{fig:compare}). We suggest that these supraparticles show, in fact, very similar catalytic activity, and the difference in their eventual efficiency is due to different mechanisms of hydrogen transport (bubbling vs.\ non-bubbling).

Second, our model predicts that catalytic systems can behave differently in the initial stages of the reaction but exhibit the same catalytic activity at long times (collapsing curves in Fig.~\ref{fig:logs}b), if they have the same values of $k_A$, $k_B$ and $\alpha$ but different values of $k_{AB}$ (Fig.~\ref{fig:theory}c,d). Indeed, in the initial stage, the reaction proceeds mostly via reduction by $\BH$, captured by $k_{AB}$. However, at later times, hydrogenation via $\Hy$ becomes dominant. The reduction rate in this region depends only on $k_A$ and the hydrogen concentration $c_{\Hy}$, which, in turn, is defined by the rates of hydrolysis $k_B$ and transport $\alpha$. The collapse, depicted in Fig.~\ref{fig:logs}b, suggests that supraparticles of Type A have the same catalytic activity with respect to hydrogenation by dissolved $\Hy$ and hydrolysis, but different catalytic activities with respect to reduction by borohydride.

Finally, the unusual activity surge observed for Type C particles is captured by the time-dependent transport $\alpha(t)$ (Eq.~\ref{eq:alpha_t}), with $\alpha_s\gg\alpha_l$ (see Fig.~\ref{fig:theory}e). When bubbling stops before full $\BH$ depletion, $\Hy$ produced by hydrolysis accumulates in the solution (see Fig.~\ref{fig:theory}f) and leads to the acceleration of $\Hy$-mediated hydrogenation of 4-NiP.

We note that the slowdown of the 4-NiP reduction over time, shown in Fig.\ref{fig:logs}a, has been observed previously, but was attributed to fractional order reaction kinetics \cite{li2013revisiting} or formation of an intermediate with very high adsorption to the catalyst \cite{gu2014kinetic,gu2015kinetic}. However, none of these models exhibit naturally the characteristic kinetic features that we have observed in our experiments: namely, the collapse of the reaction rates at long times (due to transition to $\Hy$-mediated reduction) and acceleration of the reaction upon cessation of bubbling (due to accumulation of dissolved $\Hy$). Interestingly, we do find these kinetic features in the experimental data we extracted from Refs.~\cite{li2013revisiting, gu2014kinetic}, suggesting that the data reported there is compatible with our model (see Supplementary Material).

\begin{figure}[t]

    \centering
\includegraphics[width=0.45\textwidth]{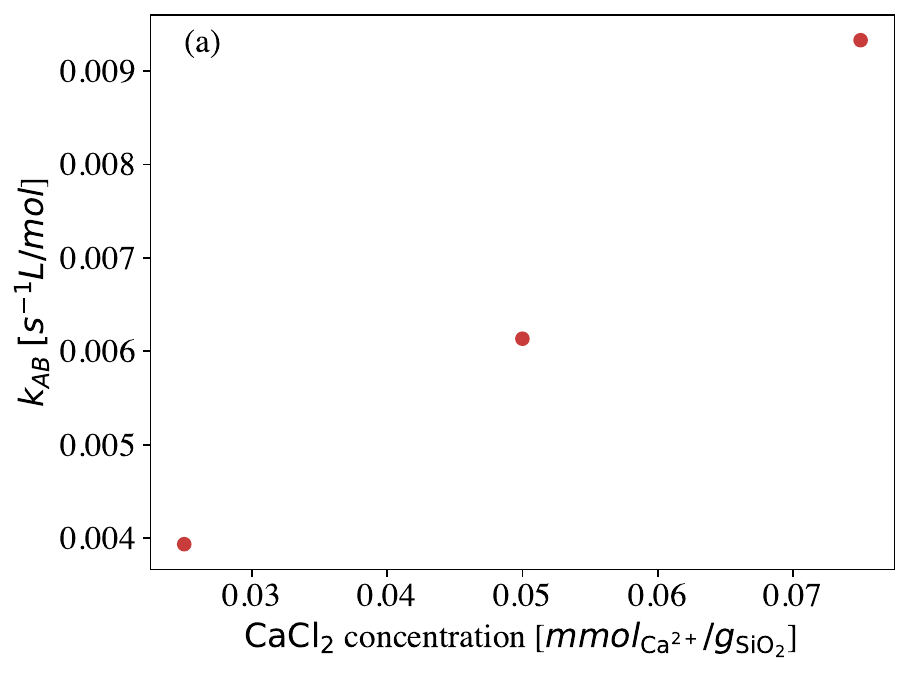}
\includegraphics[width=0.45\textwidth]{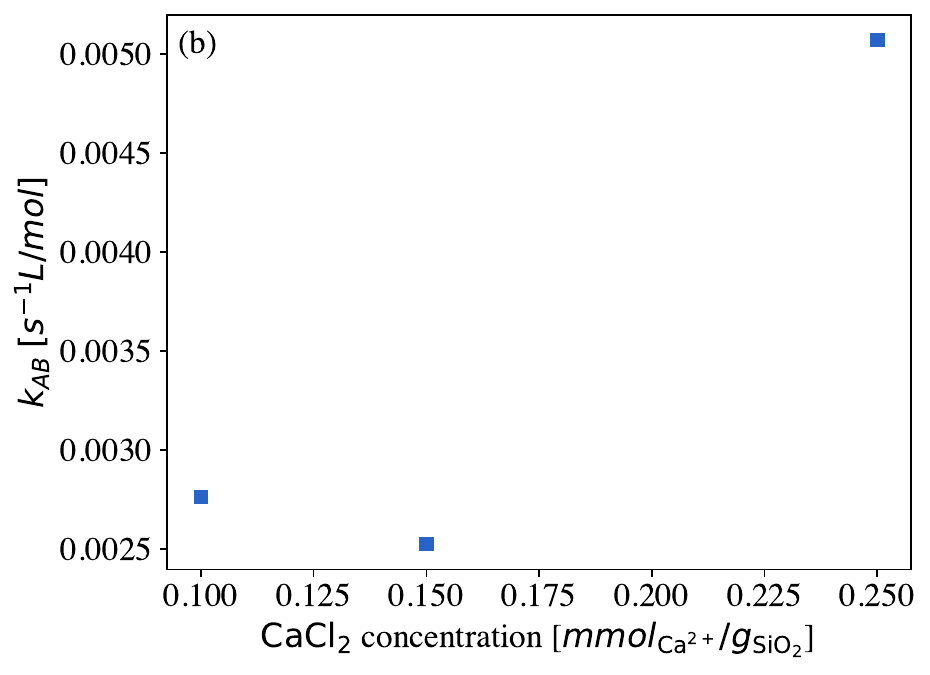}
   \caption{Fitted values of $k_{AB}$ for supraparticles made of 182 nm (a) and 19 nm $\SiO$ supraparticles (b) versus salt concentration used during fabrication. Fitting performed using the model Eq.(\ref{eq:frank_nip}-\ref{eq:frank_h2}) with parameters defined by Eq.\eqref{eq:pars_def}.}. 
    \label{fig:kAB182}
\end{figure}
Given these observations, we can fit all the data with Eqs.~(\ref{eq:frank_nip}--\ref{eq:frank_h2},\ref{eq:alpha_t}) using minimal subsets of fitting parameters, guided by the kinetic features described above.  
For Type A supraparticles, all the curves in Fig.~\ref{fig:logs}a can be fitted with the same values for $k_A$, $k_B$, $\alpha_s$, $\alpha_l$ by solely varying $k_{AB}$. 
We use the following set of common parameters:

\begin{align}
k_B=7.5\cdot10^{-3}s^{-1},&\,k_A=1.45\cdot 10^{-7}\,s^{-1},\nonumber\\
\alpha_l=1.25 \cdot 10^{-3}\,s^{-1},&\,\alpha_s=4.8\cdot 10^{-3}\,s^{-1},\label{eq:pars_def}
\end{align}
and  the following  values of bubbling time, based on experimental observations:
\begin{eqnarray}
 \text{Type A:}\;t_{bub}= 7\,\Min,\\
\text{Type B (non-bubbling):}\;t_{bub}= 0\,\Min.
\end{eqnarray}

Indeed, the collapse of the reaction rate curves at long times (Fig.\ref{fig:logs}a) suggests that $k_A$ ($\Hy$-mediated hydrogenation rate) and $\alpha_l$ ($\Hy$ transport rate in non-bubbling regime) are the same for all curves, and the fact that all curves collapse onto a master curve after the same characteristic time suggests that $k_B$ (hydrolysis rate) and $\alpha_s$ ($\Hy$ transport rate in bubbling regime) are the same as well. While other sets of parameters (with different values of $k_A$, $k_B$, $\alpha_s$, $\alpha_l$ for each curve) can also reproduce the data, the observed collapse is unlikely to be coincidental, making a shared set of parameters the most natural explanation.
Interestingly, it is possible to fit the data for supraparticles of Type B (non-bubbling), using the same set of shared parameters ($k_A$, $k_B$ and $\alpha_l$) as for Type A supraparticles; however, these fits are not unique because no clear regime change is observed, making multiple mechanisms compatible with the data (see Fig.~S1 in the Supplementary Material).

We remark that the value of $\alpha_l$, obtained by fitting the tails of the concentration curves, falls well into the expected region for the degassing rate from a stirred beaker: $\alpha_l =5\cdot 10^{-4}\ldots5\cdot 10^{-3}~\mathrm{s^{-1}},$
where we used $k_L=10^{-5}\ldots10^{-4}~\mathrm{m\cdot \sec^{-1}}$ experimentally measured for similar conditions \cite{cents2005validation,hamborg2010absorption} and $a=50~\mathrm{m}^{-1}$ (for 15 mL of the reaction mixture in a 25 mL stirred beaker).

The fitted values of the binary reaction coefficient  $k_{AB}$ are provided in Fig. \ref{fig:kAB182}. For particles of Type A $k_{AB}$ increases with the amount of salt added during supraparticle fabrication. This is in line with the fact that the average size of Pt agglomerates decreases with the increase of $\CaCl$ concentration \cite{groppe2024catalyst}, exposing a larger surface area of Pt. However, the collapse of the experimental reduction rate curves at longer times suggests that the amount of catalyst available for $\Hy$-mediated hydrogenation of $\NiP$ is the same for all supraparticles. Otherwise we would expect different reaction rates, $k_A$, for different salt concentrations.

These experimental data are compatible with the assumption that the morphology of the Pt-aggregates mostly affects the surface reaction between $\NiP$ and borohydride ions $\BH$ (i.e., $k_{AB}$), while $\Hy$-mediated hydrogenation (i.e., $k_A$) and hydrolysis  (i.e., $k_B$) are much less affected. The latter is not so surprising: 
while the initial steps of multistep hydrolysis reaction are slow at high pH, the subsequent steps are much faster and can proceed in the bulk independently of the catalyst \cite{mesmer1962hydrolysis,grzeschik2020overlooked,wang1972kinetic}. This may explain the lack of sensitivity of $k_B$ on the morphology of the Pt-aggregates that we obtain from our analysis.

For what concerns $k_A$, we recall that 4-NiP has a relatively high adsorption constant to Pt \cite{wunder2010kinetic,grzeschik2020overlooked}, which leads to almost full occupancy of catalysts by $\NiP$, which indeed is the cause of the pseudo zeroth order kinetics of its reaction with $H_2$. Accordingly, even small amounts of $\NiP$ can lead to high occupancy levels of the catalyst, irrespective of the morphology of the Pt-aggregates.

In contrast, borohydride ions have a much smaller adsorption constant \cite{wunder2010kinetic}, making the transfer hydrogenation rate very sensitive to the local concentration of borohydride, which is affected by the morphology of the aggregates and the ionic strength of the solution. 

For supraparticles of Type B, the fitted hydrogenation rate constants $k_{AB}$ are close to those for Type A, ensuring similar conversion rates at the initial time (see Fig.\ref{fig:kAB182}). The drastic difference in long-term behavior between the two types of particles can be captured by changing only one parameter: the hydrogen transport rate in the initial stage of the reaction: $\alpha_s$ for Type A and $\alpha_l$ for Type~B. However, we stress again that the datasets for particles of Type~B are compatible with different reaction mechanisms (see Fig.~S1 in Supplementary Materials).

For Type C supraparticles, the collapse of the reaction curves in the initial stage (Fig.~\ref{fig:logs}e) suggests similar activity with respect to 4-NiP reduction by borohydride. Accordingly, it is possible to fit these curves with the same $k_{AB}$, $k_B$, $\alpha_s$, $\alpha_l$ and different $t_{bub}$:
\begin{align}
k_B=5\cdot 10^{-3}\,s^{-1},&\;k_{AB}=4.33\cdot10^{-3}\,s^{-1}\text{L/mol}, \\
\alpha_l=1.25\cdot 10^{-3}s^{-1},&\,\alpha_s=0.125\,s^{-1},\label{eq:pars_def_C}
\end{align}
with bubbling times estimated from experiment:
\begin{align}
    \text{15 ALD: }t_{bub}=8 \Min\, \\
    \text{30 ALD: }t_{bub}=6 \Min.
\end{align}
 The fitted values of hydrogenation rate coefficients for 15 and 30 ALD cycles are avery close: $k_A=1.0\cdot 10^{-5}s^{-1}$ for particles fabricated with 15 ALD cycles and $k_A=1.1\cdot 10^{-5}s^{-1}$ for particles fabricated with 30 ALD cycles. Similar fits, however, could be obtained by fixing $k_A$ and varying $k_B$ or by varying both. 

The fits are provided as solid curves in (Fig.~\ref{fig:logs}a--f), with the bottom panels showing finite-difference derivatives of the fits on the experimental time grid. We can see that all the fits are in excellent agreement with the data. Note that the surge of activity observed for Type C particles (Fig.\ref{fig:logs}e,f) is reproduced solely by changing the transport regime from bubbling to non-bubbling at the time deduced from the experiment (visual disappearance of bubbles). To verify our hypothesis and assess the role of hydrogen as reducing agent, we performed an additional experiment with Type C particles fabricated with 15 ALD cycles, in which gaseous $H_2$ was bubbled through a needle submerged in the reaction mixture. While the procedure does not allow to control the actual concentration of dissolved $H_2$, we can clearly see that $H_2$-mediated reduction proceeds at a rate comparable to borohydride-driven reduction  (see magenta curve in Fig.\ref{fig:logs}e,f).

\section*{Conclusion}

By revisiting experimental data on $\Pt$-$\SiO$ supraparticles with different pore architectures (Types A, B, and C), we suggest that the reduction of $\NiP$ by $\BH$ is not governed by a single apparent rate constant but by the interplay of two reaction mechanisms: surface reaction between 4-NiP and borohydride ions on Pt nanoparticles and 4-NiP reduction by $H_2$ produced in the course of borohydride hydrolysis, which is largely controlled by the conditions of hydrogen transport. At early times, reduction by $\BH$ dominates, while at later times, hydrogenation by dissolved $\Hy$ becomes increasingly relevant once $\BH$ is depleted. Which pathway prevails is determined largely by how fast $\Hy$ is transported out of the system. 

The kinetic features demonstrated by supraparticles of Type A,B,C illustrate the combined effect of two reduction mechanisms and hydrogen transport: bubbling accelerates $\Hy$ loss and leads to incomplete conversion (Fig.~\ref{fig:logs}a, Fig.~\ref{fig:theory}a), non-bubbling systems sustain pseudo-first-order kinetics and achieve higher conversion (Fig.~\ref{fig:logs}c,  Fig.~\ref{fig:theory}c), and in some cases catalytic activity surges once bubbling ceases and dissolved $\Hy$ accumulates (Fig.~\ref{fig:logs}e,f and Fig.~\ref{fig:theory}e,f). A minimal kinetic model combining the two reduction pathways with time-dependent $\Hy$ transport reproduces these observations. The presence of the $\Hy$-mediated reduction mechanism has been confirmed in an independent experiment, in which 4-NiP reduction was achieved by bubbling gaseous hydrogen through the reaction mixture. Moreover, this mechanism can explain previously reported data for which the $\NiP$ reduction kinetics deviated from the pseudo first order model and alternative mechanisms such as fractional reaction order~\cite{li2013revisiting} or the presence of intermediates~\cite{gu2015kinetic} have been proposed. Interestingly, as shown in the Suppl. Mat., these data sets can be easily understood as the outcome of the interplay between the direct $\NiP$ reduction via $\BH$ and $\Hy$-mediated reduction.

This has important implications for using $\NiP$ reduction as a benchmark reaction. First, catalyst performance can only be compared across studies if the hydrogen-transport mechanism is the same; bubbling and non-bubbling systems should not be directly compared. In fact, our analysis shows that supraparticles with essentially identical catalytic activity (same $k_A, k_B, k_{AB}$) can exhibit very different long-term behavior depending on the rate of hydrogen transport (Fig.~\ref{fig:compare}, Fig.~\ref{fig:theory}a). Second, it is essential to track how the transport regime evolves during the reaction. If bubbling stops before $\BH$ is fully depleted, the concentration of dissolved $\Hy$ rises and triggers a sudden increase in activity (Fig.~\ref{fig:logs}e, Fig.~\ref{fig:theory}e). 
Therefore, in order to disentangle transport effects from intrinsic catalytic activity in benchmarking experiments it is important to report the hydrogen transport regime (bubbling/non-bubbling) in publications and track its changes throughout the experiment. In non-bubbling regime experiments should be reproduced at different surface-to-volume ratios (e.g. by scaling up the volume of the reaction mixture or by performing the reaction in beakers of different diameters) to exclude the dependency of the results on hydrogen transport rate. Finally, performing additional experiments on direct 4-NiP hydrogenation with $\Hy$ can help to estimate the importance of the $\Hy$-mediated mechanism for a particular catalytic system.

In addition, our analysis illustrates that $\NiP$ reduction is not a single reaction but a combination of 4-NiP reduction by borohydride, borohydride hydrolysis, and reduction by dissolved $\Hy$. Catalytic systems may therefore show different activities for each of these reactions. This is evident for Type A supraparticles with the same $\Pt$ loading but different aggregate sizes: they exhibit the same activity for reduction by dissolved $\Hy$ (see Fig.~2b, long times) but different activities for reduction by borohydride (see Fig.2b, short times).  The balance between the reduction mechanisms also depends on the chemical nature of the catalyst. For example, for silver nanoparticles 4-NiP reduction proceeds almost exclusively via binary reaction with $\BH$ at the surface of the catalyst (mechanism I in Fig.\ref{fig:sketch}), because dissolved hydrogen does not adsorb on Ag easily \cite{varshney2023competition}. In contrast, platinum is very active in catalysing both hydrolysis of borohydride and $\Hy$-mediated hydrogenation (mechanism II in Fig.\ref{fig:sketch})\cite{varshney2023competition}, but is also more prone to losses of hydrogen due to bubbling. Interestingly, the highest catalytic activity is often achieved for bimetallic nanoparticles (Ag-Pt\cite{varshney2020remarkable}, Ni-Pt\cite{ghosh2004bimetallic}, etc.\cite{pozun2013systematic}), which possibly provide an optimal balance between the two mechanisms. A model of 4-NiP reduction as a combination of two reduction mechanisms and hydrogen transport provides a convenient tool to study such systems.

Beyond 4-NiP reduction, our analysis has a broader impact: in generic reduction reactions driven by hydrogen donors, the apparent kinetics can be strongly influenced by hydrogen transport. For example, a dual reduction mechanism similar to the one we described has been reported for formic acid and related donors, where both direct hydride transfer and H$_2$ production take place~\cite{berry2019expanded,nie2021review,zhang2024cth,yu1998fa,javaid2013flow}. This suggests that the transport effects identified here are not confined to a model system, but are representative of a wider class of transfer hydrogenation reactions. Mechanistic analysis of catalytic performance should therefore account for the fate of in situ generated $\Hy$, since it can alter observed kinetics even when the underlying chemistry remains unchanged. Finally, transport effects are expected to play an important role in continuous-flow reactors~\cite{elhadad2025safety,iben2022continuous,park2025simultaneous,lomonosov2022solvent,acsomega2024nipreview}.

\section*{Conflict of interest}
The authors declare no conflicts of interest.

\section*{Acknowledgments}
We acknowledge funding by the Deutsche Forschungsgemeinschaft (DFG, German Research Foundation)—Project No. 431791331—SFB 1452.

\bibliography{kinetics}

\end{document}



\begin{center}
    \textbf{\large Supplemental Material of\\
    Effects of Hydrogen Transport on the Kinetic Regimes of 4-Nitrophenol Reduction by Sodium Borohydride}\\
    {\large Tatiana Nizkaia, Philipp Groppe, Valentin M\"uller, Jens
Harting, Susanne Wintzheimer, and Paolo Malgaretti}
\end{center}

\setcounter{page}{1}
\newcommand{\eff}{\mathrm{eff}}
\newcommand{\bnd}{\mathrm{bound}}
\newcommand{\sat}{\mathrm{sat}}
\newcommand{\DoH}{\mathrm{DoH}}

\newcommand{\Da}{\mathrm{Da}}
\newcommand{\DaBub}{\mathrm{Da}_{H_2}^\text{bub}}
\newcommand{\DaDiff}{\mathrm{Da}_{H_2}^\text{diff}}
\newcommand{\NaBH}{\text{NaBH}_4}
\newcommand{\BHPt}{\text{BH}_4\text{-Pt}}
\newcommand{\HPt}{\text{H-Pt}}
\newcommand{\Taf}{\text{Taf}}
\newcommand{\NipPt}{\text{NiP-Pt}}

\newcommand\remove[1]{\textcolor{orange}{#1}}

\setcounter{figure}{0}
\renewcommand{\thefigure}{S\arabic{figure}}

\setcounter{equation}{0}
\renewcommand{\theequation}{S\arabic{equation}}

\section{Assumptions and approximations of the model}

In order to minimize the number of parameters, we used the following simplifications in deriving the model, described in Eqs.(4a-4c) of the main text. 

\textit{First}, we use an effective pseudo zeroth-order reaction scheme for $H_2$ mediated hydrogenation of 4-NiP at all times. The effective pseudo zeroth order kinetics observed in the experiments \cite{vaidya2003synthesis,grzeschik2020overlooked} supposedly follows from the microscopic binding-unbinding kinetics of 4-NiP to the Pt nanoparticles and the reaction of the adsorbed species with  dissolved hydrogen (Eley-Rideal mechanism~\cite{eley1938heterogeneous}):
\begin{equation}
    \dfrac{dC_{\NiP}}{dt}=k_A \theta_\NiP C_{H_2},
\end{equation}
where $k_A$ is proportional to the available surface of the catalyst and $\theta_\NiP$ is the fraction of the catalyst occupied with $\NiP$:
\begin{equation}
    \theta_\NiP=\dfrac{K_\NiP C_\NiP}{1+K_\NiP C_\NiP},
\end{equation}
where $K_\NiP$ is the adsorption equilibrium constant of $\NiP$. When $K_\NiP C_{\NiP}\gg 1$, $\theta_\NiP\approx 1$ and the kinetics is pseudo zero-the order in $\NiP$. This behavior is expected to break at small densities of 4-NiP, where the full equation for coverage has to be taken into account.

\textit{Second}, we do not explicitly use Langmuir-Hinshelwood model \cite{langmuir1928hinshelwood} (accounting for the competition of the reactants for the active sites at the catalysts) to describe the surface reaction between borohydride and 4-NiP. In the literature \cite{wunder2010kinetic,gu2014kinetic,ayad2020kinetic} Langmuir-Hinshelwood approach is used to fit the dependence of the short term apparent reaction rate $k_{app}$ on initial concentrations. We use instead 
a simple binary reaction term with an \textit{effective} rate coefficient~$k_{AB}$, which can also depend on the initial concentrations of the reactants. 

\textit{Third},  we have disregarded all back-reactions, since our focus is mainly on the intermediate stages of the time evolution of the 4-NiP concentration, rather than on the eventual equilibrium state or on the initial short term effects. In particular, we disregard here the induction period, associated with the back-reaction of 4-aminophenol with dissolved oxygen \cite{wu2022spherical,menumerov2016catalytic,strachan20204,neal2019induction}, because in our experiments this induction period was shorter than the time resolution of measurements. This back reaction might also be important at late stages of the reduction process and at large surface to volume ratios, when dissolved oxygen, adsorbed from the atmosphere, starts to dominate over hydrogen. 

\textit{Finally,} we adopt a linear dependence of the hydrogen evaporation flux on the dissolved hydrogen concentration $C_{\mathrm{H_2}}$, neglecting the saturation concentration of H$_2$. This approximation is justified for evaporation from an open, stirred beaker into the atmosphere, since hydrogen is much lighter than air and does not accumulate near the gas-liquid interface. However, when an oversaturated solution is in direct contact with growing hydrogen bubbles, the effective saturation concentration depends on the bubble size through the Laplace pressure, breaking the linear dependence of te flux on $C_{H_2}$. Consequently, while the coefficient $\alpha_l$ corresponds to the degassing rate from a stirred beaker and can be estimated using gas-liquid mass-transfer models \cite{danckwerts1970gas}, the coefficient $\alpha_s$ should be regarded as an empirical parameter.

\section{Data fitting procedure }
We fit the data presented in using the theoretical model. Not surprisingly, having 4 fitting parameters for each experimental curve allows us to fit all the experimental data with high accuracy. However, is it to be expected that upon changing the concentration of CaCl$_2$ \textit{all} the rate constants have to change? 

First, we can expect that the hydrogen transport rate $\alpha$ in the non-bubbling regime should be the same for all supraparticles, because it is associated with the rate of $H_2$ evaporation from the reactor. Indeed, we can extract $\alpha$ by fitting the experimental data for $t>7$ min. For both types of particles an excellent fit is obtained with $\alpha=0.075$ $\Min^{-1}$. Second, the collapse of the experimental data for $t>7$ min (see Fig.~4b in the main text) implies that both the hydrolysis of borohydride and the hydrogenation of $\NiP$ via dissolved $\Hy$ occur at very similar rates for the particles of this set i.e. $k_A$ and $k_B$ should be the same for all Type A supraparticles. 


Interestingly, it is indeed possible to fit all the data for particles of Type A and Type B with the same values of $k_A,k_B$, $\alpha_s$, $\alpha_l$ and $t_{bub}$ taken from the experiment, by solely varying $k_{AB}$. We use the following set of common parameters:
\begin{align}
k_B=0.0075\,s^{-1},&\,k_A=1.45\cdot 10^{-7}\, s^{-1},\nonumber\\
\alpha_l=1.25 \cdot 10^{-3}\,s^{-1},&\,\alpha_s=4.8\cdot 10^{-3}\,s^{-1},\label{eq:pars_def}
\end{align}
and  the following  values of bubbling time, based on experimental observations:
\begin{eqnarray}
 \text{Type A:}\;t_{bub}= 7\,\Min,\\
\text{Type B (non-bubbling):}\;t_{bub}= 0\,\Min.
\end{eqnarray}
The values of $k_{AB}$ are fitted for each curve individually. 

\textbf{\begin{figure*}
    \centering
     \includegraphics[width=0.95\textwidth]{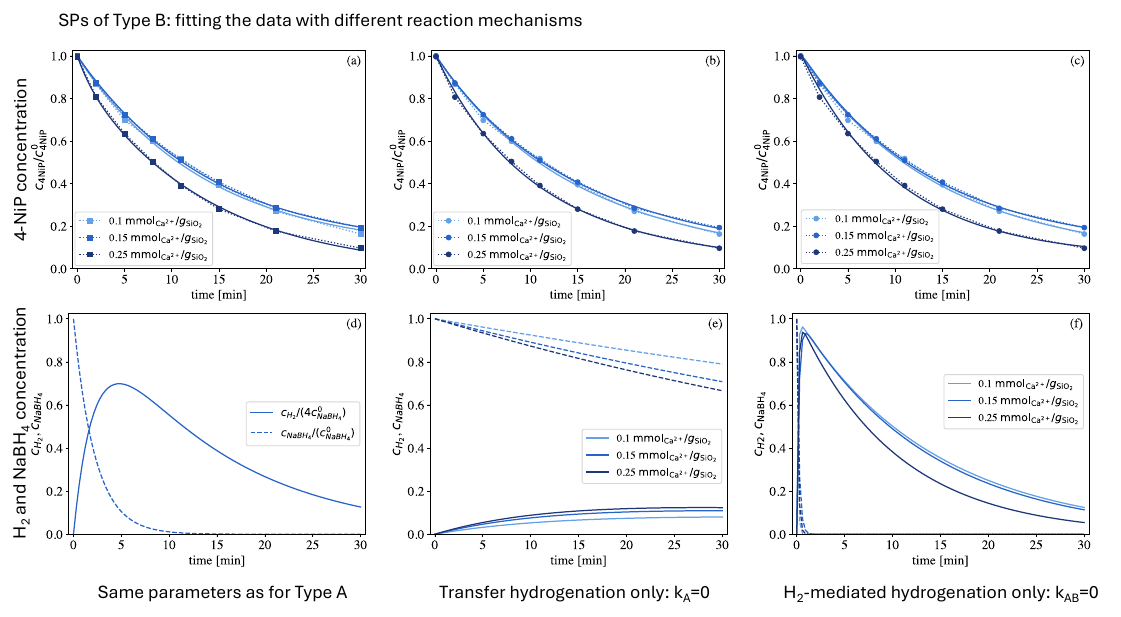}
   \caption{4-NiP concentration (top) and modeled $H_2$ concentration (bottom) for particles with 19 nm $\SiO$ grains, fabricated with different concentrations of $\CaCl$. Solid lines are fits using Eqs.~(4a-4c) with the following sets of parameters: (a,d) $k_A,\,k_B,\alpha$ defined by Eq.\eqref{eq:pars_def} and varying $k_{AB}$ (same parameters as for 182 nm) (b,e) $k_A=0$, $\alpha=1.25\cdot 10^{-3} \,s^{-1}$ and $k_{AB}$, $k_B$ fitted for each curve (hydrogenation only by surface reaction with borohydride), (c,f) $k_{AB}=0$ and fitted $k_{A}$, $k_B$, $\alpha$ (hydrogenation only by dissolved $H_2$).}
    \label{fig:c19_cNiP}
\end{figure*}
}

We remark, however, that the set of fitting parameters chosen above is not unique. Indeed, while the long-term collapse of the reaction rate curves for Type A particles imposes clear restrictions on the reaction rate coefficients, the data for supraparticles of Type B is compatible with different scenarios. As shown in the top panels of Fig.~\ref{fig:c19_cNiP}, the same data can be fitted successfully using different assumptions. For example, the fit shown in Fig.~\ref{fig:c19_cNiP}a accounts for both chemical pathways at the same time and uses the same values $k_A$, $k_B$, $\alpha$ as for Type A particles and varying $k_{AB}$. The fit in Fig.\ref{fig:c19_cNiP}b accounts only for transfer hydrogenation of 4-NiP by $\NaBH$ with $k_A=0$, the same values of $k_B$ for all curves and varying $k_{AB}$. Finally, Fig.~\ref{fig:c19_cNiP}c shows a fit assuming only $\Hy$-mediated hydrogenation of 4-NiP, with $k_{AB}=0$ and varying $k_A$, $k_B$, $\alpha$. 
Therefore, in this case it is not possible to identify the relative importance of the two chemical pathways by studying only the evolution of 4-NiP concentration. However, while the resulting $c_{\NiP}$ curves are very similar, the time evolution of the $\Hy$ concentration in the solution is different, as shown in the corresponding bottom panels in Fig.~\ref{fig:c19_cNiP}.
The bottom panels of Fig.~\ref{fig:c19_cNiP} show that the full model accounting for both chemical pathways leads to a time evolution of the $\Hy$ concentration on a time scale, similar to what was observed for the $H_2$ flux in experiments on concurrent hydrolysis and 4-NiP hydrogenation (see Fig. 1b in \cite{varshney2023competition}). At variance, accounting for solely one of the two chemical pathways leads to either monotonic growth of the concentration of $\Hy$ (Fig.\ref{fig:c19_cNiP}e)\footnote{We remark that the fit in Fig.\ref{fig:c19_cNiP}b does not fix the value of $\alpha$. Hence, in order to estimate the time evolution of $\Hy$ in  Fig.\ref{fig:c19_cNiP}e we used the same value of $k_\alpha$ that we used for Type A supraparticles and Fig.\ref{fig:c19_cNiP}a.}, or a very steep surge of $\Hy$ concentration (Fig.\ref{fig:c19_cNiP}f), followed by slow relaxation.

Therefore, our model suggests a route to identify the underlying chemical pathway by performing simple additional measurements. In fact, since in our model   the outflow of hydrogen is proportional to its concentration, measuring of $\Hy$ flux in the experimental condition may differentiate between the different reaction mechanisms. Note, that the measurements of hydrogen flux do not need to be very precise: it is sufficient to estimate whether the flux increases, decreases or experiences a non-monotonous evolution in time.

While different sets of parameters may be used to fit the experimental curves with the model in Eqs.~(4a-4c), we find the one proposed in Eq.\eqref{eq:pars_def}  particularly interesting since it allows us to fit all experimental curves upon varying a single parameter $k_{AB}$.

\section{Analysis of literature data}
In this section we revisit previously published data on 4-nitrophenol reduction, focusing on the studies where deviations from pseudo-first-order kinetics have been reported. The datasets have been extracted from the figures of corresponding papers using the online tool Automeris v5 \cite{Automeris}.

\begin{figure*}
    \centering
     \includegraphics[width=0.45\textwidth]{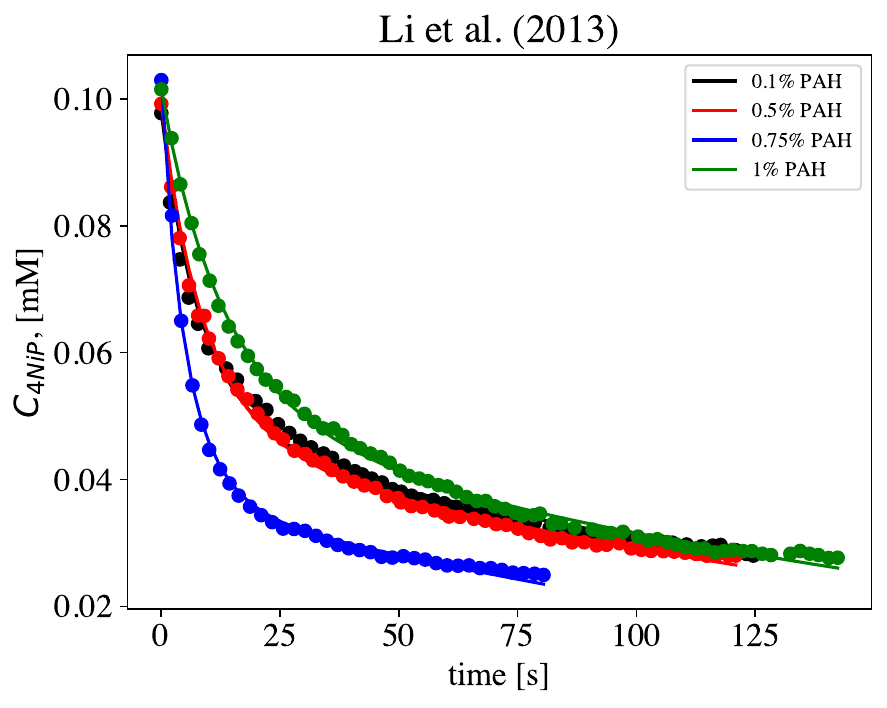}
     \includegraphics[width=0.45\textwidth]{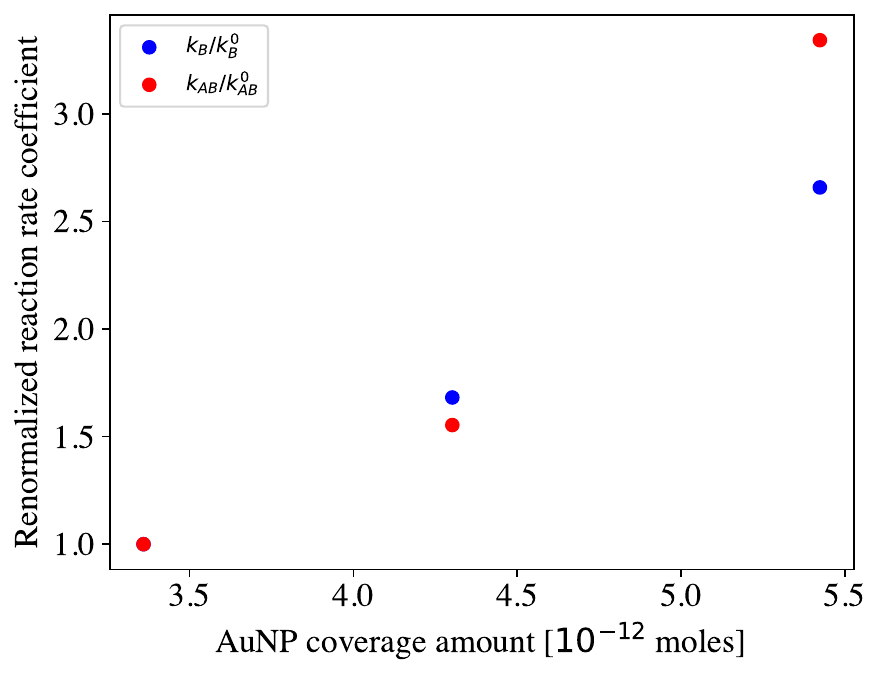}
   \caption{4-NiP concentration (left) and fitted reaction rates (right) for 4-NiP reduction on Au-NP embedded onto polymeric spheres. Data taken from \cite{li2013revisiting} is represented by circles, solid curves are fits using our model Eq.(4a-4c). }
    \label{fig:li}
\end{figure*}

\begin{figure*}
    \centering
     \includegraphics[width=0.45\textwidth]{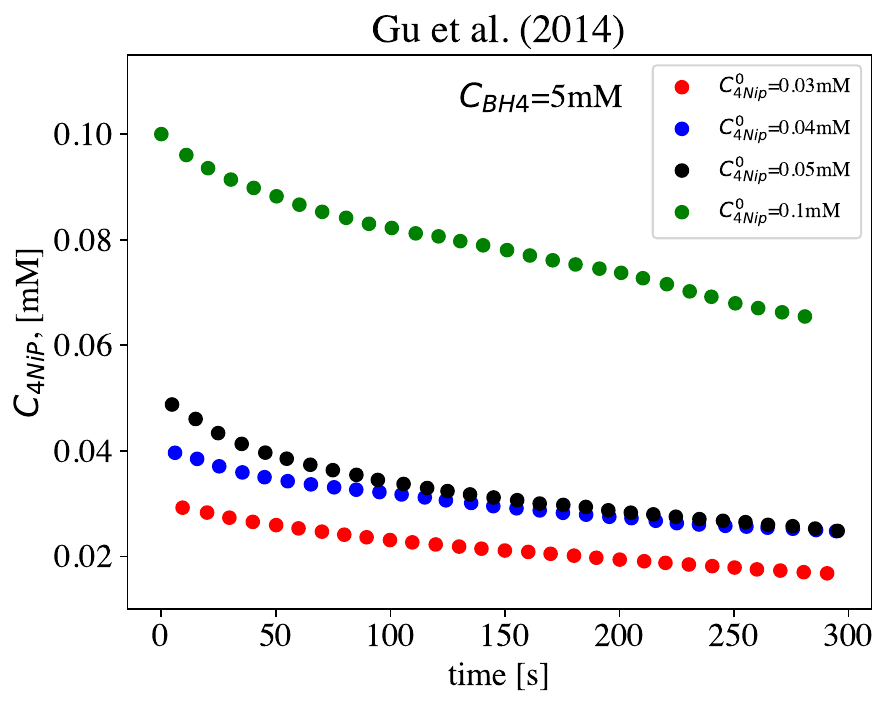}
     \includegraphics[width=0.47\textwidth]{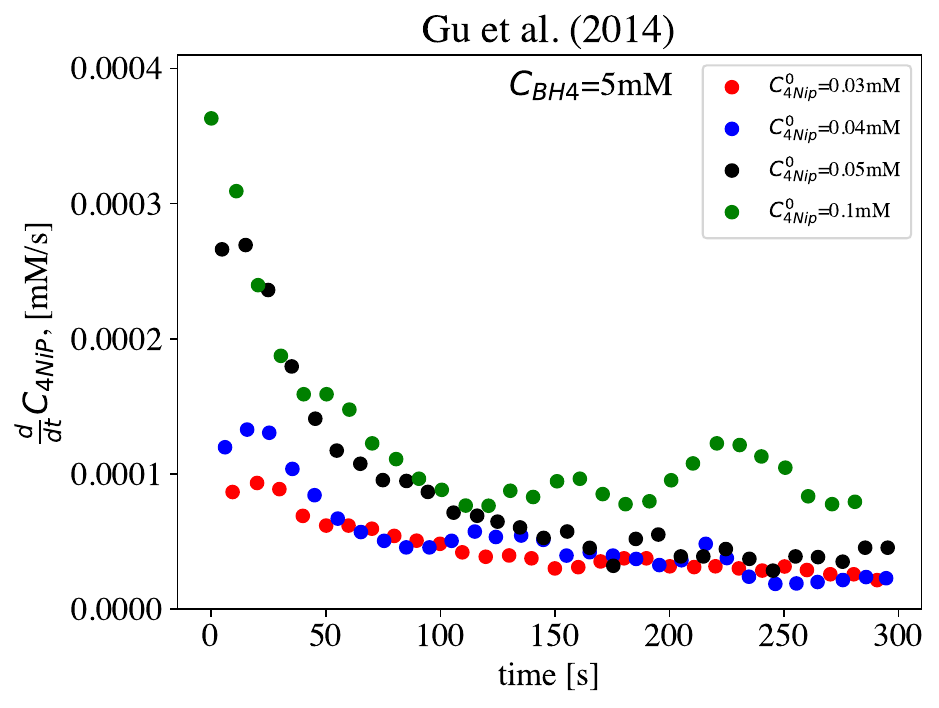}\\
   \caption{4-NiP concentration (left) and its time derivative (right) for different initial concentrations of 4-NiP. Data taken from Ref.\cite{gu2015kinetic}, Fig.3.}
    \label{fig:gu}
\end{figure*}

We start by revisiting work by Li et al.\cite{li2013revisiting}, in which Au nanoparticles supported by polymeric sub-microspheres were used as the catalyst. In this work authors observed significant deviations of 4-NiP reduction kinetics from pseudo-first order and reported formation of $H_2$ bubbles. To explain the data the authors proposed a fractional order kinetic model with respect to 4-NiP and found that the reaction order depends on the surface coverage of Au particles. However, we note that the apparent fractional order kinetics can also result from an overlap of two different kinetic regimes operating at similar time scales. To illustrate the point, we took the concentration curves from Fig.3 in \cite{li2013revisiting} and fitted them using our two-mechanisms model Eq.(4a-4c)(solid curves in Fig. \ref{fig:li}a). We can see that the apparent fractional order arises naturally when both hydrolysis and hydrogenation rates increase with increasing Au-NP coverage (see fitted parameters $k_B$, $k_{AB}$ provided in Fig.\ref{fig:li}b). Interestingly, a similar system with silver nanoparticles on a resin support \cite{kanti2019temperature} did not show any deviations from pseudo first order kinetics despite the reported presence of $\Hy$ bubbles. This is in line with the fact that Ag nanoparticles do not catalyse $\Hy$-mediated hydrgogenation, because dissociative adsorprion of hydrogen on silver is very weak \cite{varshney2023competition}. Therefore only mechanism I (binary reaction between $\NiP$ and $\BH$ on the surface of the catalyst) with first-order reaction kinetics is present.

Kinetics with different apparent reaction rates at short and long times has been also observed and studied in \cite{gu2014kinetic,gu2015kinetic}. To explain this behavior authors proposed a reaction scheme involving formation of an intermediate species (4-hydroxylaminophenol) that has a very strong adsorption to Pt and blocks the active sites causing the slowdown of the reaction. 
Our model provides an simple alternative explanation to this behavior: a shift from direct reaction with borohydride to hydrogenation via dissolved $\Hy$. To illustrate our point, we have extracted the data from Fig.3 in \cite{gu2015kinetic}, featuring 4-NiP reduction by $\NaBH$ on $\mathrm{SPB-Au}_{75}\mathrm{Pd}_{25}$ nanoalloys at different initial concentrations of 4-NiP and $\NaBH$. While the plots in \cite{gu2015kinetic} are provided in terms of the ratio $C_{\NiP}/C^0_{\NiP}$, we reconstruct the $\NiP$ concentration and reduction rate curves, by multiplying the data by the respective initial concentration $C_{\NiP}^0$ (see Fig.\ref{fig:gu}).  Interestingly, at large excess of borohydride $C_{\NaBH}^0/C_{\NiP}^0>100$ (red, blue and black symbols in Fig.\ref{fig:gu}) we observe the collapse of the reaction rate curves at long times, similar to what we have witnessed for Type A particles.
Such collapse is typical for the transition from borhydride-mediated to $H_2$-mediated reduction of 4-NiP. Moreover, the non-monotonous shape of the rate curve for $C_{\NiP}^0=0.1\;mM$ suggests a reaction acceleration associated with transition from bubbling to non-bubbling regime and build-up of dissolved $H_2$, similar to what was observed for Type C particles. However, since the presence or absence of bubbling has not been reported in \cite{gu2015kinetic}, it is impossible to check this hypothesis.
\bibliography{kinetics}